\documentclass[aps,prd,amssymb,showpacs]{revtex4}
\usepackage{epsfig}
\begin{document}

\title{On characteristic initial data for a star orbiting a black hole}

\author{Nigel T. Bishop${}^{1}$,
    Roberto G\'omez${}^{2,3}$,
    Luis Lehner${}^{4}$,
    Manoj Maharaj${}^{5}$ and
    Jeffrey Winicour${}^{3,6}$}
\affiliation{
${}^{1}$Department of Mathematical Sciences,
University of South Africa, P.O. Box 392, Pretoria 0003, South Africa \\
${}^{2}$Pittsburgh Supercomputing Center,
4400 Fifth Ave., Pittsburgh, Pennsylvania 15213, \\
${}^{3}$Department of Physics and Astronomy,
University of Pittsburgh, Pittsburgh, Pennsylvania 15260\\
${}^{4}$Department of Physics and Astronomy,
Louisiana State University, Baton Rouge, LA 70810,\\
${}^{5}$Department of Mathematics and Applied Mathematics,
University of Durban-Westville, Durban 4000, South Africa,\\
${}^{6}$Albert Einstein Institute, Max Planck Gesellschaft,
Am M\"uhlenberg 1, D-14476 Golm, Germany
}

\date {Nov 21st 2004}

\begin{abstract}

We take further steps in the development of the characteristic approach to
enable handling the physical problem of a compact self-gravitating object, such
as a neutron star, in close orbit around a black hole. We examine different
options for setting the initial data for this problem and, in order to shed
light on their physical relevance, we carry out short time evolution of this
data. To this end we express the matter part of the characteristic gravity code
so that the hydrodynamics are in conservation form. The resulting gravity plus
matter relativity code provides a starting point for more refined future
efforts at longer term evolution. In the present work we find that, independently
of the details of the initial gravitational data, the system quickly flushes out
spurious gravitational radiation and relaxes to a quasi-equilibrium state with
an approximate helical symmetry corresponding to the circular orbit of the
star.

\end{abstract}


\maketitle

\section{Introduction}
\label{sec:intro}

The problem of computing the evolution of a self-gravitating object, such as a
neutron star, in close orbit about a black hole is of clear astrophysical
importance, both to understand systems thought to drive such spectacular
phenomena as gamma-ray bursts (see, for instance,~\cite{rees}) and to predict
the details of the gravitational radiation which could be observed by the new
generation of  gravitational wave detectors (see, for instance,~\cite{ct}).
Furthermore, the dynamics of the system could be quite rich. For instance, the
tidal interaction  between the black hole and the star could be strong enough
to tidally disrupt the star with a consequent drastic change in the emitted
waves~\cite{vallisneri}. Alternatively,  as has recently been suggested
\cite{davies},  only a portion of the star's mass might be transferred to the
black hole and angular momentum transfer might boost what
remains of the star into a wider orbit. The system would then undergo another
inspiral phase and this process might repeat itself. Consequently, the expected
gravitational waves would display a very rich structure of several
`chirp-phases'.

A better understanding of these possibilities requires, as basic building
blocks,  solving Einstein equations coupled to an appropriate matter field
without restrictive assumptions. To this end, long-term well behaved numerical
simulations must be available. Presently, several numerical relativity  codes
for treating the two body problem are either being developed or 
planned~\cite{baumgarte,shibata,miller,hawke} , as the
know-how for simulating Einstein equations becomes more mature. Although most
of these efforts concentrate on the Cauchy approach to Einstein equations,
there are alternatives approaches which have been successful for specific
problems. In particular, the characteristic formulation of general relativity
has shown remarkable robustness to deal with single black hole space-times.
Stable axisymmetric studies of Einstein equations coupled to perfect fluids
have been achieved~\cite{siebel2,siebel1}; and our previous work has produced
stable three-dimensional  characteristic  numerical relativity codes for vacuum
space-times~\cite{hpn} and for fluid space-times with very small
pressure~\cite{mat}.

This paper develops further methodology necessary for the
evolution of a self-gravitating star or other object in orbit near
a Schwarzschild black hole. Towards this final goal, we examine
the issue of setting initial data. In either the characteristic or
Cauchy approaches to this problem, a  serious source of physical
ambiguity is the presence of spurious gravitational radiation in
the initial gravitational data. Because the characteristic
approach is based upon a retarded time foliation, the resulting
spurious outgoing waves can be computed by carrying out a short
time evolution. We carry out such a study in the present work. We
find that, independently of the details of the initial
gravitational data, such spurious waves quickly radiate away, and
that the system relaxes to a quasi-equilibrium state with an
approximate helical symmetry corresponding to the circular orbit
of the star. This result provides important physical justification
of recent approaches for initializing the Cauchy problem which are
based on imposing an initial helical
symmetry~\cite{klein,shibatafriedman,yocook,grandclement,andrade}.

We also examine two useful tools which can be applied to long term
simulations as well as to the initial data problem: (i) a tool to monitor the
development of a singularity in the null coordinates and (ii) a tool to use
co-rotating coordinates so that an orbiting star remains approximately at a
fixed coordinate position.

A crucial ingredient of any hydrodynamical simulation is the proper handling of
shocks and discontinuities and the proper conservation of baryonic mass.
Robust numerical techniques are available to this end, which express the system
in a conservation form designed for handling discontinuities in the
fluid~\cite{leveque}. In particular, high resolution shock capturing schemes
utilize the fluid's characteristic propagation speeds to capture the
discontinuities in an accurate way. In the present work, we follow the
formalism presented in~\cite{philiptoni}, where the general relativistic
hydrodynamic equations are presented in a conservation form adapted to both the
Cauchy and the characteristic formulations. High resolution shock capturing
schemes have been successfully incorporated in the characteristic  approach in
the case of axisymmetric space times~\cite{siebel2,siebel1}. Although this is
the ideal  approach to the hydrodynamic problem, here we implement a less  time
consuming algorithm due to  Davis~\cite{davis}. This approach has limitations in dealing
with shock formation but is sufficient for our present purpose to study the
initial data problem.

The initial data problem for the characteristic formulation of
relativity~\cite{hpn} has received little attention in comparison with that for
the Cauchy problem. This is partly because the data for the characteristic
formulation is unconstrained, whereas the data for the Cauchy formulation has
to satisfy an elliptic system of constraints. Even so, in both formulations,
care must be taken to set matter data that represents the intended physical
problem. In particular, it is not trivial to set gravitational data free of
spurious gravitational radiation. A method for initializing characteristic data
based upon a correspondence with Newtonian
theory~\cite{newt1,newt2} guarantees that the resulting gravitational
radiation obeys the Einstein quadrupole formula in the Newtonian
regime~\cite{iww,qf}. However, as discussed in~\cite{newt1}, the domain of
applicability of the method is limited to when (i) the Newtonian potential
$\Phi$ of the star satisfies $\Phi \ll 1$ and (ii) relative velocities are much
slower than the speed of light. The second issue could be easily dealt with by
making the orbital radius $a$ of the star large compared to the black hole mass
$M$. However, the following practical reasons make it desirable to initialize
the star in a close orbit with $a \le 10M$ (for which $v \ge c/3$):

\begin{itemize}

\item Since the orbital period of the star around the black hole scales as
$a^{3/2}$,  a simulation starting from large $a$ would require very time
consuming runs in a regime that could be described approximately by less
expensive perturbative treatments.

\item When using spherical coordinates, the resolution at large $a$ worsens
when using a uniform grid. Computer resources limit the number of angular
grid points that can be used. Thus, adequate resolution of the star requires
that $a$ be small.

\item The most interesting physics, intractable by other means, occurs when the
compact object is at small $a$.

\end{itemize}

Construction of mathematically consistent initial data is much simpler than
construction of physically meaningful data. While the former involves the
solution of constraint equations (if any), the latter is less mathematically
explicit. Not only does one want matter data describing an orbiting star but
also gravitational data with minimal spurious radiation. Success must be
gauged through actual evolution of the data and analysis of the resulting
space-time. \footnote{Some notion of the proximity to stationarity on the initial
hypersurface can be gained without evolution by the approach introduced in
\cite{dain}.}

There are basically no constraint equations for the gravitational initial data
in the characteristic formulation so that mathematically consistent data is
trivial to prescribe. In order to investigate the physical problem we proceed
by investigating gravitational data based upon two completely different
underlying assumptions:

\begin{itemize}

\item Gravitational initial data obtained by means of the
Newtonian correspondence method

\item  Spherically symmetric gravitational data corresponding
to a shear-free initial null hypersurface.

\end{itemize}

The first method applies only to a quasi-Newtonian regime.
The second method ignores the focusing effect of the star which introduces
shear in its nearby null rays. Nevertheless, in both cases the gravitational
field relaxes to a state of approximate helical symmetry after about a
light-crossing time for the system. These results support the view that the
details of the initial gravitational data may not be important as long as they
are reasonable and some time is evolved in which the spurious radiation flushes out.

The plan of the paper is as follows: in Sec.~\ref{sec:nota} with present a summary of previous
work on the characteristic formulation of numerical relativity and on
conservative hydrodynamics. The construction of initial matter and
gravitational data is covered in Sec~\ref{sec:newt} in both stationary and
co-rotating frames. In Sec.~\ref{sec:caustic} we discuss the conditions under
which coordinate singularities might develop. Some details of the numerical
implementation are presented in Sec.~\ref{s-num}. Computational tests and
results are given in Sec.~\ref{sec:tests}.

\section{Summary of previous results and notation}
\label{sec:nota}

\subsection{Characteristic formulation of Einstein equations}

The formalism for the numerical evolution of Einstein's equations, in null cone
coordinates, is well known~\cite{rai83,hpn,cce} (see
also~\cite{ntb93,ntb90,rai83}), and is based upon the analytic treatments
of~\cite{bondi,sachs,tam}. For the sake of completeness,
we give here a summary of the formalism, including some of the necessary equations.
The version of the gravity code being used here is fully described
in~\cite{roberto,particle}. It has most recently been applied in~\cite{modemode}.

We use coordinates based upon a family of outgoing null hypersurfaces.
We let $u$ label these hypersurfaces, $x^A$ $(A=2,3)$, label
the null rays and $r$ be a surface area coordinate. In the resulting
$x^\alpha=(u,r,x^A)$ coordinates, the metric takes the Bondi-Sachs
form~\cite{bondi,sachs}
\begin{eqnarray}
   ds^2 & = & -\left(e^{2\beta}(1 + {W \over r}) -r^2h_{AB}U^AU^B\right)du^2
        -2e^{2\beta}dudr -2r^2 h_{AB}U^Bdudx^A \nonumber \\
        & + & r^2h_{AB}dx^Adx^B,
\label{eq:bmet}
\end{eqnarray}
where $h^{AB}h_{BC}=\delta^A_C$ and
$det(h_{AB})=det(q_{AB})$, with $q_{AB}$ a unit sphere metric.
We work in stereographic coordinates $x^A=(q,p)$ for which the unit sphere
metric is
\begin{equation}
q_{AB} dx^A dx^B = \frac{4}{F^2}(dq^2+dp^2),
\mbox{ where }
        F=1+q^2+p^2.
\end{equation}
Our previous work used $P=1+q^2+p^2$, here we change notation to $F$
because we will use $P$ to represent pressure.
We also introduce a complex dyad $q^A=\frac{F}{2}(1,i)$
with $i=\sqrt{-1}$. For an arbitrary Bondi-Sachs metric,
$h_{AB}$ can then be represented by its dyad component
\begin{equation}
J=h_{AB}q^Aq^B/2,
\end{equation}
with the spherically symmetric case characterized by $J=0$.
We introduce the (complex differential) eth operators $\eth$ and $\bar \eth$
~\cite{eth}, as well as a number of auxiliary variables
$K=h_{AB}q^A \bar q^B /2$, $U=U^Aq_A$, $Q_A=r^2 e^{-2\,\beta}h_{AB}U^B_{,r}$,
$Q=Q_Aq^A$, $B=\eth\beta$, $\nu=\bar\eth J$ and $k=\eth K$.

The Einstein equations $G_{ab}=0$ decompose into hypersurface equations,
evolution equations and conservation laws. Here we  note that the hypersurface
equations form a hierarchical set for $\nu_{,r}$, $k_{,r}$, $\beta_{,r}$,
$B_{,r}$, $Q_{,r}$, $U_{,r}$ and $W_{,r}$, and the evolution equation is an
expression for $(rJ)_{,ur}$. The explicit form of the equations is given
in~\cite{roberto} in the vacuum case and in~\cite{mat} for the matter terms.
Note the following correction to the matter source term in Eq.~\cite{mat}--(31)
\begin{eqnarray}
    && 2 \left(rJ\right)_{,ur}
    - \left((1+r^{-1}W)\left(rJ\right)_{,r}\right)_{,r} =
    -r^{-1} \left(r^2\eth U\right)_{,r}
    + 2 r^{-1} e^{\beta} \eth^2 e^{\beta}- \left(r^{-1} W \right)_{,r} J
    + N_J \nonumber \\
    && +\frac{4 e^{2\beta}\pi(\rho + P)}{r}
    \left( (J \bar{V}_{ang} - K V_{ang})^2 +V^2_{ang}   \right),
\label{eq:Ju}
\end{eqnarray}
where $V_{ang}=v_A q^A$~\cite{mat} (with $v_\alpha$ the velocity) and $N_J$ are
defined in~\cite{hpn} and~\cite{roberto}. The remaining Einstein equations are
the conservation conditions which are satisfied here because of the simple
choice of boundary data.

The null cone problem is normally formulated in the region of space-time
between  a timelike or null worldtube $\Gamma$ and ${\cal I}^+$. We represent
${\cal I}^+$ on a finite grid by using a  compactified radial coordinate
$x=r/(1+r)$. The numerical grid is regular in $(x,q,p)$, and consists of two
stereographic patches covering the north and south hemispheres, each containing
$n_x \times n_q \times n_p$ grid points. The $x-$grid covers the range $[0.5, 1]$, and each
angular grid patch extends at least two grid points beyond the equator so
that there is an overlap region.

The mass of the Schwarzschild black hole is denoted as $M$, and in all the
computational tests we will take $M=1$. The star has mass
$m$ and radius $R_*$.

\subsection{Characteristic hydrodynamics in conservation form}
\label{sec:hydro}

The integration of the hydrodynamical equations is done in a more accurate way
if the system can be expressed in conservation form. This allows for better
conservation of baryonic mass and the possibility of exploiting the
characteristic structure of the equations to resolve shocks via Godunov
methods~\cite{leveque}. Here we use the formalism developed in
~\cite{philiptoni}, where the equations $\nabla_a J^a=0$, $\nabla_a T^{ab}=$
are expressed in conservation form as,
\begin{equation}
\partial_{x^{0}} (\sqrt{-g} {\bf U}^{A})
+ \partial_{x^{j}} (\sqrt{-g} {\bf F}^{jA})
= {\bf S}^{A} \, ,
\label{eq:cons-law}
\end{equation}
where $U^A=(D,S^i,E)$ are ``conservative'' variables defined as,
\begin{eqnarray}
\label{eq:D}
D     & = & {\bf U}^{0} = J^{0}  = \rho u^{0} \, ,  \\
S^{i} & = & {\bf U}^{i} = T^{0i} = \rho h u^{0} u^{i} + P g^{0i} \, ,  \\
E     & = & {\bf U}^{4} = T^{00} = \rho h u^{0} u^{0} + P g^{00} \, .
\label{eq:E}
\end{eqnarray}
and the fluxes and sources are
\begin{eqnarray}
{\bf F}^{j0} & = &  J^{j}  = \rho u^{j} \, , \nonumber \\
{\bf F}^{ji} & = &  T^{ji} = \rho h u^{i} u^{j} + P g^{ij} \, , \nonumber \\
{\bf F}^{j4} & = &  T^{j0} = \rho h u^{0} u^{j} + P g^{0j} \, ,
\label{eq:fluxes} \\ \nonumber
\\
{\bf S}^{0} & = & 0 \, , \nonumber \\
{\bf S}^{i} & = & - \sqrt{-g} \, \Gamma^{i}_{\mu\lambda} T^{\mu\lambda}
\, , \nonumber \\
\label{eq:sources}
{\bf S}^{4} & = & - \sqrt{-g} \, \Gamma^{0}_{\mu\lambda} T^{\mu\lambda} \, .
\end{eqnarray}
The state of the system is uniquely described in terms of the geometry, the
primitive matter variables $(\rho,P,h,u^a)$ or the conservative variables, the
fluid's equation of state and the normalization condition $u^a u_a = -1$. In
the particular case of a perfect fluid which we consider here, the relation between primitive and
conservative variables is straightforward and does not require expensive
inversion methods (see \cite{philiptoni} for details).

\section{Initial and boundary data}
\label{sec:newt}

There exist only rough physical guidelines for prescribing initial matter and
gravitational field data in general relativity for a star orbiting a black
hole. Here we present initial data which is at least mathematically consistent
with Einstein's equations and has some underlying connection with the Newtonian
picture of a star in equilibrium, which is orbiting a collapsing central object.
The Newtonian picture is not a good approximation to the relativistic regime in
which we run our simulations. Nevertheless, the justification for this approach
is that the evolutions, presented in Sec.~\ref{sec:tests}, relax on the
order of a light crossing time to a more astrophysically realistic state, as
extraneous gravitational wave content in the initial data is radiated away, and
the gravitational field adapts to the approximate symmetry of the matter
distribution. This relaxed state is remarkably independent of the initial
gravitational data.

\subsection{The initial matter data}
\label{sec:phys}

We prescribe the initial matter data
within a simple Newtonian framework. We take the star
to be a spherically symmetric polytrope of index $n=1$~\cite{chandra1} with
\begin{equation}
       \rho=\frac{m \sin{\frac{R \pi}{R_*}}}{4 R R_*^2}
\end{equation}
for $R \le R_*$, and $\rho=0$ for $R > R_*$, where $\rho$ is the density at
radius $R$. Denoting the pressure by $P$, the equation of state is
\begin{equation}
         P=\frac{2 R_*^2 \rho^2}{\pi}.
\end{equation}
Note that the maximum value of $P/\rho$ is $m/2R_*$, so that Newtonian theory
gives a good approximation to an equilibrium configuration provided the
polytrope is not near its Schwarzschild radius. We prescribe the initial matter
velocity to be uniform across the polytrope. In the evolutions considered later
the center of the polytrope will be placed at $(q=p=0,r=a)$, with the velocity
set for a circular orbit, so that $V^r=V^p=0$ initially.

The assumption that the polytrope is spherically symmetric is always valid
mathematically since the initial data is freely specifiable, but the star
remains in equilibrium only if there is no tidal force. Of course, it is known
how to compute the tidal distortion of a polytrope in equilibrium within both
Newtonian gravity and post-Newtonian general relativity -- see for example~\cite{uryu,lom}. However, in the simulations of a star orbiting a black
hole presented in this paper, the polytrope is not stable against tidal
disruption, i.e. the simple Roche indicator $\epsilon_R \equiv
R_*/a\sqrt[3]{2M/m} \gg 1$. Thus, for the present work, there is no point in
calculating an equilibrium configuration for the tidal distortion.

In order to obtain physical characteristic initial data for the problem
we transform the above Newtonian initial data into density, velocity
and gravitational fields within the framework of a Bondi-Sachs metric in
general relativity. As discussed in Sec.~\ref{sec:intro}, this needs to be done in
a relativistic regime in which the Newtonian correspondence
method~\cite{newt1,newt2,iww} is only roughly approximate.

The density and velocity matter fields are regarded as being given in a
Lorentzian frame in the Minkowski space-time in which the center of mass of the
polytrope is at rest. Then a coordinate transformation is made from this local
Lorentz frame to the global Bondi-Sachs coordinates. This approach is strictly
valid only if space-time curvature can be neglected in the vicinity of the star,
which requires that the radius of the star must be small and that its
gravitational source effect must be negligible. The approach has a proper
Newtonian correspondence in the limiting case of small velocities, small
$m/R_*$ and large $a$.

\subsubsection{The density and velocity fields}
\label{sec:sr}

We prescribe initial data ($\rho$, $V^\alpha$) for the localized distribution
of matter described in Sec.~\ref{sec:phys}. The matter is described in a
(locally) inertial frame $S_I=(t,x,y,z)$ in which the center of mass of the
matter is instantaneously at rest at the origin. Globally, we have Bondi-Sachs
coordinates centered about the black hole (with mass $M$). To the extent that
the gravitational effect of the matter can be neglected, the geometry is
spherically symmetric and can be described in Eddington-Finkelstein coordinates
$S_E=(u,r,q,p)$. The origin of $S_I$ is at $u=0,r=a,q=p=0$ and, for simplicity,
we choose a central matter velocity of the form $(V^u,0,V^q,0)$ in $S_E$.
Except in the case $M=0$, the curvatures of $S_I$ and $S_E$ are different, and
thus it is not possible to construct a unique global transformation between
$S_I$ and $S_E$. However, we can construct a transformation that is valid near
the origin of $S_I$.

We proceed by constructing a locally inertial frame $S^\prime=
(t^\prime,x^\prime,y^\prime,z^\prime)$ whose origin is at
$O=(u=0,r=a,q=0,p=0)$  and with the $S^\prime$ axes pointing in the $(u,r,q,p)$
directions. In general, $S_I$ is moving relative to $S^\prime$. When the black
hole mass $M=0$, the space-time is flat and there is an unambiguous
transformation from $S^\prime$ to $S_E$
\begin{eqnarray}
t^\prime & = &  u + (r-a)  \nonumber \\
x^\prime & = & \frac{2 r q}{1+q^2+p^2}  \nonumber \\
y^\prime & = & \frac{2 r p}{1+q^2+p^2}  \nonumber \\
z^\prime & = & \frac{2 r}{1+q^2+p^2}- r-a.
\label{eq:trs'm}
\end{eqnarray}
When $M\ne 0$, the metric is Schwarzschild in the  Eddington-Finkelstein form
\begin{equation}
ds^2=-\Big( 1-\frac{2M}{r}\Big)du^2 -2 du \; dr
      +\frac{4 r^2}{(1+q^2+p^2)^2}(dq^2+dp^2).
\label{eq:EFm}
\end{equation}
An orthonormal tetrad aligned with $S^\prime$ and defined in
$S_E$ at the point $O$ is given by
\begin{eqnarray}
X_{(t)}^a & = & (\Big( 1-\frac{2M}{a}\Big)^{-\frac{1}{2}},0,0,0) \nonumber \\
X_{(x)}^a & = & (0,0,\frac{1}{2a},0) \nonumber \\
X_{(y)}^a & = & (0,0,0,\frac{1}{2a}) \nonumber \\
X_{(z)}^a & = & (-\Big( 1-\frac{2M}{a}\Big)^{-\frac{1}{2}},
           \Big( 1-\frac{2M}{a}\Big)^{\frac{1}{2}},0,0).
\end{eqnarray}
In a neighborhood of $O$, $S_E$ and $S^\prime$ are related by
the transformation
\begin{eqnarray}
u & = & \Big( 1-\frac{2M}{a}\Big)^{-\frac{1}{2}} t^\prime
    -\Big( 1-\frac{2M}{a}\Big)^{-\frac{1}{2}} z^\prime  \nonumber \\
r-a & = &  \Big( 1-\frac{2M}{a}\Big)^{\frac{1}{2}} z^\prime \nonumber \\
q & = &   \frac{x^\prime}{2a} \nonumber \\
p & = &   \frac{y^\prime}{2a},
\end{eqnarray}
from which we construct the inverse transformation
\begin{eqnarray}
t^\prime & = & \Big( 1-\frac{2M}{a}\Big)^{\frac{1}{2}} u
    +\Big( 1-\frac{2M}{a}\Big)^{-\frac{1}{2}} (r-a)  \nonumber \\
x^\prime & = & 2 a q  \nonumber \\
y^\prime & = & 2 a p  \nonumber \\
z^\prime & = & \Big( 1-\frac{2M}{a}\Big)^{-\frac{1}{2}} (r-a).
\label{eq:trs'ef}
\end{eqnarray}
In the $M=0$ case, the transformation is unambiguous and global, but when $M
\ne 0$ the transformation to a locally inertial system can only be defined in a
neighborhood of $O$. Nevertheless we need a transformation that is valid in
a region around $O$, which we construct by combining (\ref{eq:trs'm}) and
(\ref{eq:trs'ef}) to obtain
\begin{eqnarray}
t^\prime & = & \Big( 1-\frac{2M}{r}\Big)^{\frac{1}{2}} u
           +\Big( 1-\frac{2M}{r}\Big)^{-\frac{1}{2}} (r-a)  \nonumber \\
x^\prime & = & \frac{2 r q}{1+q^2+p^2}  \nonumber \\
y^\prime & = & \frac{2 r p}{1+q^2+p^2}  \nonumber \\
z^\prime & = & \Big(\frac{2 r}{1+q^2+p^2}- r-a\Big)
           \Big( 1-\frac{2M}{r}\Big)^{-\frac{1}{2}}.
\label{eq:trs'}
\end{eqnarray}
The transformation (\ref{eq:trs'}) reduces to (\ref{eq:trs'm}) when $M=0$,
and to (\ref{eq:trs'ef}) near the point $O$.

It is most convenient to describe the matter in an inertial frame in which it
is at rest which, in general, is not the case for $S^\prime$. In order to
achieve this we need to make a Lorentz transformation to coordinates
$S_I=(t,x,y,z)$. In the case that the matter velocity is entirely in the
$q$-direction, which includes the case of a quasi-circular orbit around the
black hole, the transformation between $S_E$ and $S_I$ is
\begin{equation}
t  =  \gamma (t^\prime-v x^\prime), \;\;
x  =  \gamma (x^\prime-v t^\prime), \;\;
y  =  y^\prime, \;\;
z  =  z^\prime,
\label{eq:L}
\end{equation}
where $v$ is the velocity between $S_I$ and $S^\prime$, and $\gamma$ is the
usual Lorentz factor, $\gamma=(1-v^2)^{-\frac{1}{2}}$. Then combining
(\ref{eq:L}) and (\ref{eq:trs'ef}) we find
\begin{eqnarray}
t & = & \gamma \Bigg( \Big( 1-\frac{2M}{r}\Big)^{\frac{1}{2}} u
           +\Big( 1-\frac{2M}{r}\Big)^{-\frac{1}{2}} (r-a)
           -\frac{2 r q v}{1+q^2+p^2}
           \Bigg)      \nonumber \\
x & = & \gamma \Bigg( \frac{2 r q}{1+q^2+p^2}
           -v \Big( 1-\frac{2M}{r}\Big)^{\frac{1}{2}} u
           -v \Big( 1-\frac{2M}{r}\Big)^{-\frac{1}{2}} (r-a)
           \Bigg)         \nonumber \\
y & = & \frac{2 r p}{1+q^2+p^2}  \nonumber \\
z & = & \Big(\frac{2 r}{1+q^2+p^2}- r-a\Big)
           \Big( 1-\frac{2M}{r}\Big)^{-\frac{1}{2}}.
\label{eq:trs}
\end{eqnarray}
The transformation (\ref{eq:trs}) is sufficient for setting scalar initial
matter data, such as the density. Given an $S_E$ grid at $u=0$, we find for
each point the coordinates $(x,y,z)$. The density is then determined by its
values in $S_I$. We also need to set the initial 4-velocity. In $S_I$ the
4-velocity is $V^{(I)}_a=(-1,0,0,0)$. Thus in $S_E$ the covariant 4-velocity is
\begin{eqnarray}
V^{(E)}_a & = & \frac{\partial x^{(I)b}}{\partial x^{(E)a}} V^{(I)}_b
        =  -\frac{\partial t}{\partial x^{(E)a}}  \nonumber \\
       & = & \gamma \bigg(-\Big( 1-\frac{2M}{r}\Big)^{\frac{1}{2}},
          \nonumber \\
    & & \frac{2qv}{1+q^2+p^2}-\Big( 1-\frac{2M}{r}\Big)^{-\frac{1}{2}}
    +\frac{(r-a)M}{r^2} \Big( 1-\frac{2M}{r}\Big)^{-\frac{3}{2}},
    \nonumber \\
     & &  \frac{2rv(1+p^2-q^2)}{(1+q^2+p^2)^2}, \nonumber \\
         & &       -\frac{4vqpr}{(1+q^2+p^2)^2} \bigg).
\label{eq:VEFa}
\end{eqnarray}

In order to set a value of $v$ for a quasi-circular orbit, we need
$V^{(E)}_a$ and $V^{(E)a}$ at $O$. We find
\begin{equation}
V^{(E)}_a=\gamma \bigg(-\Big( 1-\frac{2M}{a}\Big)^{\frac{1}{2}},
-\Big( 1-\frac{2M}{a}\Big)^{-\frac{1}{2}},  2va,0 \bigg)
\end{equation}
\begin{equation}
V^{(E)a}= \gamma \bigg( \Big( 1-\frac{2M}{a}\Big)^{-\frac{1}{2}},0,
                \frac{v}{2a},0 \bigg)
\end{equation}
A circular orbit has zero radial acceleration. Thus, applying the geodesic
equation, we find
\begin{equation}
v=\sqrt{\frac{M}{a-2M}}, \; \; \gamma=\sqrt{\frac{a-2M}{a-3M}}
\label{eq:v}
\end{equation}
As expected, a circular orbit can exist only if $a>3M$.

\subsection{Initial gravitational data}
\label{sec:J}

There is not a well-developed theory for initializing the
gravitational field of a star in close orbit about a black hole. We
consider two options here. The first is to simply set
\begin{equation}
       J(u=0,r,x^A)=0.
\end{equation}
In the absence of matter, this choice would eliminate all spurious
gravitational waves but in the presence of matter it is physically unrealistic.
It implies that the null rays generating the initial null hypersurface are
shear-free, i.e. there is no bending of light by the star as would expected in
a normal astrophysical scenario. This introduces spurious gravitational waves
which superimpose with the gravitational field of the star to cancel the
bending effect. The second choice of $J$ is quasi-Newtonian gravitational
data~\cite{newt1,newt2,iww,qf} determined by a Newtonian correspondence method
which introduces the correct bending effect and radiation content in the
Newtonian limit.

This quasi-Newtonian data is astrophysically realistic only when the Newtonian
potential and the matter velocity are small. These conditions are only very
roughly satisfied for the relativistic binary considered here. The Newtonian
potential approaches unity as $r$ approaches $2M$ and the velocity of the body
approaches that of light. Nevertheless, it is reasonably simple to compute
quasi-Newtonian initial data and it is interesting to compare the resulting
evolutions with those initialized by $J=0$, e.g. to compare their
spurious radiation content and the quasi-equilibrium states to which they relax.

The procedure for determining quasi-Newtonian data involves solving a sequence
of Poisson equations which, in the Newtonian limit,
lead to exact agreement with the Einstein
quadrupole formula for the initial gravitational radiation content
in the system. Here
we apply the method only at leading order where
Eq.~(3.8) of~\cite{iww} gives (in the present notation)
\begin{equation}
     (r^2 J_{,r})_{,r} = - 2 \eth^2 \Phi
\label{eq:J}
\end{equation}
where $\Phi$ is the Newtonian potential satisfying
\begin{equation}
     \nabla^2 \Phi=4\pi \rho
\end{equation}
in terms of the Euclidean Laplacian $\nabla^2$ obtained in the Newtonian limit.
The computation of $\Phi$ is straightforward and Eq.~(\ref{eq:J}) can then be
solved by a combination of numerical and analytic techniques  subject to
boundary conditions that determine the integration constants. Equation
(\ref{eq:J}) establishes a Newtonian correspondence on the initial null
hypersurface $u=0$. The sequence of successive Poisson equations ensure that
(\ref{eq:J}) is satisfied in time in terms of a Taylor expansion in $u$.

At the $r=2M$ surface of the black hole, we choose the boundary conditions
\begin{equation}
     J=J_{,r} =0 \;\;\mbox{at }r=2M
     \label{eq:bc}
\end{equation}
for solving the Poisson equations.
For $M=0$, these boundary conditions reduce to regularity
conditions at the vertex of the outgoing null cones.
Details of the calculation are given in the Appendix.

\subsection{Inner boundary data}
\label{sec:boun}

Our aim is to simulate a star which is in orbit close to an astrophysical
object undergoing gravitational collapse. We model the exterior field of the
collapsing object by the vacuum geometry of the outgoing $u=const$ null
hypersurfaces  whose  inner boundary is a white hole horizon. In terms of the
Kruskal picture of an isolated Schwarzschild black hole this inner boundary
would be the past branch of the $r=2M$ hypersurface. Thus our simulations
correspond to the characteristic evolution of data on two intersecting null
hypersurfaces, the initial outgoing null hypersurface ${\cal N}_0$ and the
white hole horizon ${\cal H}^-$. Since the orbit of the star is exterior to
${\cal H}^-$ the inner boundary data for the matter is trivial.

The construction of general gravitational data for such a double-null initial
value problem has been reduced to the integration of propagation equations
(ODE's) along the null generators of ${\cal H}^-$~\cite{nulltube}. The free
data on  ${\cal H}^-$ consist of the specification of its conformal metric,
represented by $J$, and its {\em shift} represented by the Bondi variable
$U$.

Here we make the simple choice that $J=0$ vanishes on  ${\cal H}^-$. One of the
propagation equations (the Raychaudhuri equation) then implies that the {\em
intrinsic expansion} of ${\cal H}^-$ is constant along its generators. We
choose the initial value of this expansion to vanish so that $r=const$ on
${\cal H}^-$. We then set $r=2M$ on the sphere ${\cal S}^-$ where ${\cal H}^-$
and ${\cal N}_0$ intersect. Thus the inner boundary ${\cal H}^-$ is described
by
\begin{equation}
             J=0 \mbox{ at } r=2M.
\label{eq:JM}
\end{equation}

These choices simplify the integration of the remaining propagation equations on
${\cal H}^-$. Choosing the ``shift'' of ${\cal H}^-$ to vanish leads to the
further simplification that $U=0$ on ${\cal H}^-$. We consider this choice first.
(Below we consider the shift corresponding to coordinates co-rotating with the
orbiting star.) The remaining data at ${\cal S}^-$ is the initial {\em twist} of
the horizon, which determines $U_{,r}$ on the horizon and the initial {\em outward
expansion} of ${\cal S}^-$, which determines the initial value of $\beta$. In the
spirit of simplicity, we set the twist to zero, which assigns zero angular
momentum to the white hole, and we set $\beta$ to zero, which in the pure
Schwarzschild case sets the scale of retarded time $u$ on the horizon to the
standard inertial time measured by observers at null infinity.

Because all free initial inner boundary data at ${\cal S}_0$ have been chosen to
be identical to the inner boundary data of a Schwarzschild horizon in standard
Eddington-Finkelstein form, and because we have chosen $J=0$ on ${\cal H}^-$,
this data propagates along the generators of ${\cal H}^-$ to give the
Eddington--Finkelstein boundary values for the corresponding Bondi
variables, namely $J=U=U_{,r}=\beta=0,W=-2M$. Alternatively, we could
prescribe Minkowskian inner boundary data by setting the value of the
inward expansion to match that of a ingoing (collapsing) Minkowski light
cone.

The Bondi evolution system also requires the value of $J_{,r}$ on
${\cal H}^-$. However, for the simple boundary data presented above, the
horizon propagation equations imply that
$J_{,r}=const$ on ${\cal H}^-$, and our boundary condition (\ref{eq:bc})
then implies $J_{,r}=0$.

\subsubsection{ Co-rotating data}
\label{sec:U}

It is also desirable to be able to specify data at $r=2M$ such
that the angular coordinates of a star in uniform circular orbit
around the black hole with angular velocity $\Omega$ remain
constant. In terms of a standard angular coordinate $\theta$, this
can be arranged by the transformation $\theta\rightarrow
\theta+\Omega u$ on the data presented above. This introduces a
{\em shift} on the horizon which is represented by a non-zero
value of $U$. At large distances from the horizon this shift
becomes superluminal, but calibration tests \cite{wobble} have
shown that this does not preclude solving the characteristic
initial value problem. In the Schwarzschild case, the vector field
$\partial_u$ in the non-rotating coordinates is the time
translation Killing vector $T$, whereas in the rotating
coordinates $\partial_u$ equals the {\em helical} Killing vector
$T+\Omega\Phi$ where $\Phi$ is a rotational Killing vector.

The calculation of the required $U$, which must be carried out in the
stereographic $(q,p)$ angular coordinates used in the code, was performed by
means a computer algebra script. We start with the Bondi-Sachs metric in standard form
for a Minkowskian space-time, with coordinates $(r,q,p,u)$, and transform
$(r,q,p)$ to Cartesian coordinates $(x,y,z)$, as specified in
Eq.~(\ref{eq:trs'm}) (here the Cartesian coordinates are written
unprimed, even though they are written primed in Eq.~(\ref{eq:trs'm})).
We then perform a rigid rotation about the $y$-axis through an angle
$\theta$, leading to Cartesian coordinates $(x^\prime,y^\prime,z^\prime)$
with
\begin{eqnarray}
x^\prime & = & x \cos \theta + z \sin \theta \nonumber \\
y^\prime & = & y \nonumber \\
z^\prime & = & z \cos \theta - x \sin \theta.
\end{eqnarray}
Finally, we transform back to Bondi-Sachs coordinates
$(r^\prime,q^\prime,p^\prime,u^\prime)$ by computing the transformation
$(r,q,p,u) \rightarrow (r^\prime,q^\prime,p^\prime,u^\prime)$ and the Jacobian for the
transformation when $\theta$ is small; then these are employed to find the
metric of Minkowskian space-time in $(r^\prime,q^\prime,p^\prime,u^\prime)$
coordinates. In order to check the calculation, via a computer algebra script we checked
that all components of the Riemann tensor of the transformed metric are
zero. Next, we apply the coordinate transformation $(r,q,p,u) \rightarrow
(r^\prime,q^\prime,p^\prime,u^\prime)$ to the $S_E$ metric (\ref{eq:EFm}).
We find that the metric remains in Bondi-Sachs form with $J=\beta=0,W=-2$
and
\begin{eqnarray}
g_{qu} & = & \frac{2 f r^2 (1+q^2-p^2)}{(1+q^2+p^2)^2} \nonumber \\
g_{pu} & = & \frac{4 q p f r^2}{(1+q^2+p^2)^2},
\end{eqnarray}
where $f=d\theta/du$, and for convenience the ${}^\prime$ symbols have been
dropped. Thus,
\begin{equation}
U=- f \frac{ 1 + q^2-p^2+2 q p i}{1+q^2+p^2}.
\label{eq:Uf}
\end{equation}
The value of $f$ is set by the condition that the center of the polytrope should
be at rest with respect to the ${}^\prime$ coordinates. We achieve this by
transforming the covariant velocity (\ref{eq:VEFa}) to the
${}^\prime$ coordinates, and then evaluating the contravariant velocity at
$r=a$, $q=p=0$. We find
\begin{equation}
V^q= -\frac{\gamma}{2a\sqrt{1-\frac{2M}{a}}}
\bigg(f a - v\sqrt{1-\frac{2M}{a}}\bigg);
\end{equation}
thus $V^q=0$ if we set
\begin{equation}
f=\frac{v\sqrt{1-\frac{2M}{a}}}{a}.
\end{equation}
When $v$ is given by Eq.~(\ref{eq:v}),
\begin{equation}
f=  \sqrt{\frac{M}{a^3}}.
\end{equation}
then Eq.~(\ref{eq:Uf}) becomes
\begin{equation}
U=- \sqrt{\frac{M}{a^3}}\frac{ 1 + q^2-p^2+2 q p i}{1+q^2+p^2}.
\label{corotation}
\end{equation}
Note that $U_{,r}=0$. Furthermore, the transformation to ${}^\prime$
coordinates does not change the formulas for $V_r$, $V_q$ and $V_p$ given in
Eq.~(\ref{eq:VEFa}); the formula for $V_u$ is changed, but that is not
part of the required initial data.

\section{Regularity of the null coordinates}
\label{sec:caustic}

The Bondi description of an asymptotically flat space is based upon the
surface area coordinate $r$, which is assumed to increase monotonically to
infinity along the outgoing null rays. This monotonicity is built into the
initial data by assuming that the matter variables and the gravitational data
$J$ are well-behaved functions of $r$, for $2M\le r\le \infty$. However,
certain choices of matter data can lead to gravitational data that is
unrealistic astrophysically. For example, consider a spherical star of mass
$m$ and radius $R_*$ located at $r=a$, with $a>>2M$ so that the linear
approximation for the bending of light by the star is valid. Then, under
normal astrophysical conditions, when

\begin{equation}
             a =\frac{R_*^2} {4 m }.
\label{caustic}
\end{equation}
the outgoing null rays which graze the surface of the star would be bent by the
star so that they would travel to infinity along asymptotically parallel
trajectories. This lack of expansion of the outgoing rays would be a breakdown
of the regularity of the $r$-coordinate, and it would also introduce a large
shear in the geometry of the outgoing null rays. This loss of regularity can be
mathematically avoided by prescribing gravitational data with zero (or small)
shear along with the matter data for the star. But clearly that would not
represent an astrophysically realistic problem. Note that, for a given star
mass $m$, this breakdown occurs for large values of $a$ and not when the star
is close to $r=2M$.

It is important to monitor this effect. An affine parameter $\lambda$
measured along the radially outgoing null rays must of course increase
monotonically in an asymptotically flat space. Thus the relevant quantity
to measure is the expansion $\partial_\lambda r$, which in a Bondi
coordinate system is given by
\begin{equation}
    \partial_\lambda r =e^{-2\beta}.
\end{equation}
In our setup, $\partial_\lambda r =1$, i.e $\beta=0$, at the inner $r=2M$
boundary, but the Bondi hypersurface equation
\begin{equation}
    \partial_r \beta= \frac{r}{8} (J_{,r} \bar{J}_{,r}-K_{,r}^2)
            +2\pi r(\rho+P) v_r^2
\end{equation}
implies that $\beta$ increases outward monotonically. Thus the expansion
$\partial_\lambda r$ decreases monotonically and reaches its smallest value at
infinity. Initial data for which $\partial_\lambda r |_\infty \approx 1/10$,
corresponding to $\beta_\infty \approx 1$, would either represent an extreme
astrophysical scenario (such as a star about to enter a black hole) or would
signal an imminent breakdown of the $r$-coordinate.

In Sec.~\ref{sec:mexpans} we illustrate how this effect can be monitored by
measuring $\beta_\infty$. This is important in order to avoid wasting
computational time trying to simulate systems which are either unrealistic
astrophysically, or where a coordinate singularity will develop.

\section{Numerical Implementation}
\label{s-num}

\subsection{Hydrodynamical equations}

We implement an algorithm of Davis~\cite{davis}, which can be
regarded as adding artificial dissipation where needed to a
MacCormack scheme, by means of a slope delimiter procedure.
To do this requires knowledge of the largest of the eigenspeeds (speed of characteristic
modes of the principal part of the fluid's evolution equations) in each
direction. The expressions for these can be read-off from those worked
out in~\cite{philiptoni}.

This dissipation, shown in $1D$ for simplicity, takes the form
\begin{equation}
U^{n+1}_i = \tilde U^{n+1}_{i} + (D_{i+1/2}-D_{i-1/2}) \, ;
\end{equation}
where $\tilde U^{n+1}_{i}$ is the update from the MacCormack step and
$D_{i+1/2}$ is defined as,
\begin{equation}
D_{i+1/2} = \left( K_{j+1/2}^+(r^+_j)
+  K_{j+1/2}^-(r^-_{j+1}) \right ) \left ( U^n_{j+1} - U^n_j \right )
\end{equation}
with
\begin{eqnarray}
r^{\pm}_j &=& \frac{<\Delta U^n_{j-1/2},
\Delta U^n_{j+1/2}>}{< \Delta U^n_{j\pm 1/2}, \Delta U^n_{j\pm 1/2}>} \\
K^{\pm}_{j+1/2} &=& \frac{1}{2} C(v) \left ( 1-\Psi (r^{\pm}_{j+1/2}) \right ) \, .
\end{eqnarray}
The values of $v$, $C$ and $\Psi$ are
\begin{eqnarray*}
v &=& \max \{ |\lambda_j| \} \frac{dt}{dx}
\end{eqnarray*}
\begin{eqnarray*}
C(v) = \left\{ \begin{array}{ll}
        v (1-v) & \mbox{if $v\leq 0.5$}  \\
        0.25 & \mbox{otherwise \, ,}
               \end{array}
        \right.
\end{eqnarray*}
\begin{eqnarray*}
\Psi(r) = \left\{ \begin{array}{ll}
        \min(2r,1) & \mbox{if $r \geq 0$}  \\
        0 & \mbox{otherwise \, .}
               \end{array}
        \right.
\end{eqnarray*}

\subsection{Update strategy}

In this work we combine the PITT characteristic vacuum code  developed in
\cite{cce,hpn,roberto} (and thoroughly tested and applied in a variety of situations,
see for instance \cite{mat,modemode,wobble,fission}) with the general
relativistic hydrodynamic equations provided above. Although this is the first
application of the combined equations in three-dimensional settings, we follow
closely the strategy pursued successfully in two-dimensional  scenarios
\cite{siebel1,siebel2}. The hierarchy of integration of the equations is
basically the following:

\begin{enumerate}
\item With data on an initial hypersurface ${\cal N}_{u}$, the metric is updated
to the new level  ${\cal N}_{u+\Delta u}$. Here the fluid variables at the
intermediate level ${\cal N}_{u+\Delta u/2}$, needed for the integration of  the
metric equations, are approximated, to first order in time, by their values at ${\cal
N}_{u}$. Note that since the typical propagation speeds of the fluid is less than
the speed of light, this approximation in general is acceptable.
\item Next, the general relativistic hydrodynamic equations are updated to ${\cal N}_{u+\Delta u}$.
\end{enumerate}

Although the integrations of the GR/hydrodynamical equations are
written in second order form when the fluid/GR variables are frozen, the above
procedure is formally only a first order accurate approximation.
Formally higher order schemes can be obtained by iterating several times per time-step,
where ``intermediate'' values of the fields are employed (consisting of the average of field values at
${\cal N}_{u}$ and the previously obtained fields at ${\cal N}_{u+\Delta u}$.
Alternatively, a more efficient scheme can be devised by keeping an extra level
of the fluid variables (at ${\cal N}_{u-\Delta u}$), so that one can extrapolate
to second order their values at  ${\cal N}_{u+\Delta u/2}$. In the present
work, since we concentrate in rather short time evolutions,  we have
opted not to do this in order to reduce the amount of time required by
the code. Additionally, during the time evolutions we use a $\Gamma$-law
equation of state given by $p=(\Gamma-1)\rho \epsilon$, with $\Gamma=1+1/n$ and
ignore the effects of viscosity and magnetic fields, since their dynamical time
scales are much longer than those considered here.

\section{Computational tests}
\label{sec:tests}

Our main goal here is to study the initial evolution phase of a star in orbit
around a black hole as a preliminary step toward carrying out long term
evolutions. Of primary concern is whether the system can be initialized in a
way which allows a meaningful simulation of the ensuing inspiral and capture of
the companion star.  The strategy is to begin with a rough choice of
gravitational data, and then show that the system quickly relaxes to a state
which provides physically reasonable matter and gravitational data, which can
in turn be used to initialize a longer evolution.

Several options are available for carrying out this study in a discriminating
and efficient way. They involve the choice of initial data; the choice of
coordinates fixed with respect to the black hole or co-rotating with the
orbiting star;
the possibility of setting the black
hole mass to zero; and the use of switches in the code that allow running the
gravitational field or the hydrodynamics in a frozen mode (which expedites the
turn-around of the numerical tests). In all tests, the star is initialized with
a uniform angular velocity about the black hole corresponding to the circular
orbital velocity of its center. We then choose from the following specific
options:

\begin{itemize}

\item Initial gravitational data given by either $J=0$ or quasi-Newtonian data.

\item Full evolution or evolution with the internal hydrodynamics of the
star frozen, i.e. the gravitational field reacts to a rigidly orbiting
source.

\item Fixed coordinates ($U=0$ at the inner boundary) or co-rotating coordinates
($U$ given by the value calculated in Sec.~\ref{sec:U}).

\item Inner boundary consisting of either a mass $M=1$ Schwarzschild event horizon or of an ingoing null cone in Minkowski space-time (no black hole).

\end{itemize}

We also carry out experiments to check the range of stellar mass, size
and orbital radius for which light bending effects do not introduce
singularities in the null coordinates. The observed behavior of the code is
described in the sub-sections below. Unless otherwise indicated, all norms refer to the $L_2$ norm.

\subsection{Code convergence}

The convergence of the PITT code has been established in a series of
papers for evolution in vacuum or with fluids having negligible
pressure~\cite{hpn,wobble,modemode,mat}. The new ingredient here is the addition
of a fluid obeying
a polytropic equation of state. The fluid is treated numerically by a first
order accurate algorithm. To illustrate this we place the star in a space-time without a
black hole and monitor the solution under different grid resolutions.

\begin{figure}
\begin{center}
\epsfig{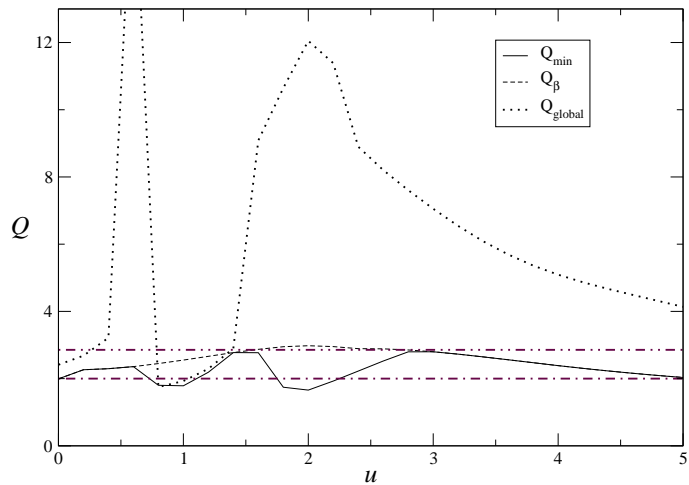}
\end{center}
\caption{For the case $m=10^{-5}$, $M=0$, $R_*=3$, $a=9$ and
grid sizes given by $45^2 \times 63$, $65^2 \times 93$ and
$85^2 \times 123$, the different self-convergence factors are obtained.
The behavior observed is consistent with at least first order convergence.}
\label{convergence1}
\end{figure}

Because the evolution of the hydrodynamics with the present serial code
is considerably time consuming, the convergence test must be based upon
limited grid resolution.
We evolved the case $m=10^{-5}$, $R_*=3$, $a=9$ using
quasi-Newtonian initial gravitational data.
Three different grid sizes were used: $45^2 \times 63$;
$65^2 \times 93$ and $85^2 \times 123$ (stereographic $\times$ radial). With this
choice of grid sizes, the middle and higher resolution grid spacings correspond
to $2/3$ and $1/2$ the grid spacing of the base grid, respectively. In order
to assess the convergence behavior of the implementation we examine the
following quantities,
\begin{eqnarray}
Q_{\beta} &=&  \frac{ |\beta(\Delta)-|\beta(\Delta 2/3)|}{|\beta(\Delta 2/3 )-|\beta(\Delta /2)|}  \\
Q_{min} &=& min \left \{ \frac{ |F_I(\Delta)-|F_I(\Delta 2/3)|}{|F_I(\Delta 2/3 )-|F_I(\Delta /2)|} \right \} \\
Q_{global} &=& \sqrt{ \frac{ \Sigma_{Ij} \left ( F_{Ij}(\Delta)- F_{Ij}(\Delta 2/3) \right )^2}
{\Sigma_{Ij} \left ( F_{Ij}(\Delta 2/3)- F_{Ij}(\Delta/2 ) \right
)^2} } \, .
\end{eqnarray}
The capital index $I$ ranges over all gravitational fields, i.e. $F_I = \{
\beta, J, U, W\}$, and the index $j$ over all points in the base grid which are
common to the middle and high resolution grid. The factors $Q_{\beta}$,
$Q_{min}$ and $Q_{global}$ measure the convergence of the field $\beta$, the
minimum of the convergence rate obtained for all gravitational field
quantities $Q_{min}$ and a global value $Q_{global}$ including all fields, respectively.  The
results are displayed in figure \ref{convergence1}, where the horizontal lines
provide the value for first and second order convergence. The observed
behavior indicates a convergence rate consistent with first order convergence.

\subsection{Evolution for different gravitational initial data}

Independently of the initial data $J$, the star settles into
``quasi-equilibrium'' after a couple of hydrodynamical times given by $2R_*$,
i.e.  all variables relax to roughly constant values with respect to a
co-rotating observer. Of the small remaining variation, a major portion
subsequently dies off after a couple of crossing times $2a$, the time needed for
the star to communicate with the inner boundary and back. This latter behavior
is observed whether or not the internal hydrodynamics of the star is evolved.
After this relaxation period, the time dependence of the gravitational field is
mainly determined by the motion of the star.

\begin{figure}
\begin{center}
\epsfig{file=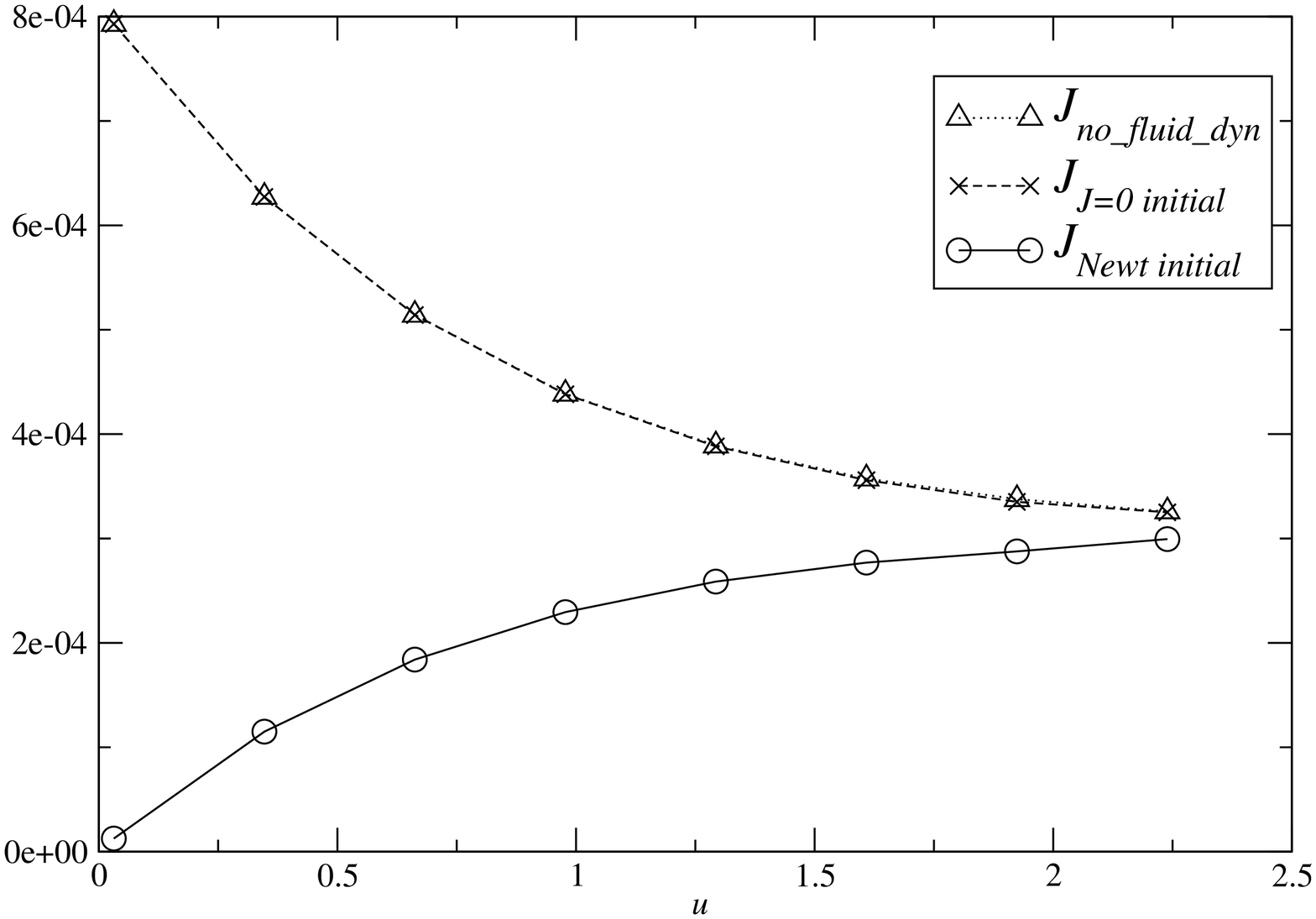,height=3.in,angle=0}
\end{center}
\caption{Behavior of $||J||$ for the case with no black hole.
Three scenarios are plotted: Full evolution (hydrodynamics + gravity) using
initial data $J=0$ (dashed line with ``x'' symbols); full evolution using
quasi-Newtonian initial data  (solid line with circular symbols); and
gravitational evolution with frozen internal hydrodynamics using
quasi-Newtonian data (dotted lines with triangular symbols). After some
transient behavior, lasting until about $u\sim 2$, all curves roughly approach
the same value.}

\label{mink_comp}
\end{figure}

\begin{figure}
\begin{center}
\epsfig{file=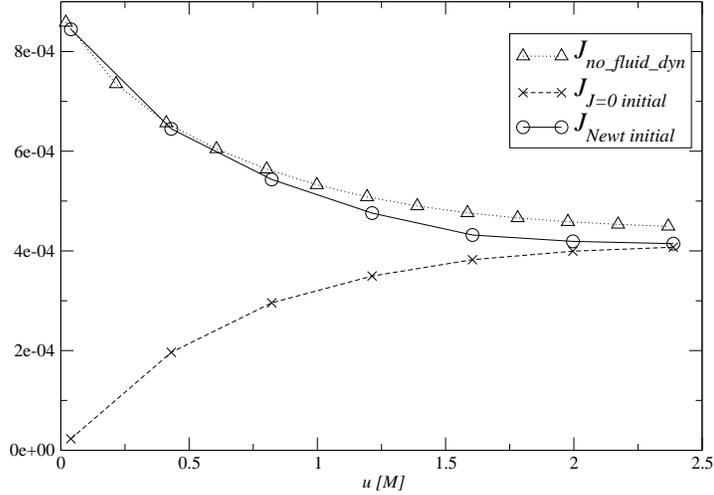,height=3.in,angle=0}
\end{center}
\caption{Behavior of $||J||$ for the case of an $M=1$ black hole. Again
three scenarios are plotted: Full evolution using initial data $J=0$ (dashed
line with ``x'' symbols); full evolution using quasi-Newtonian  initial data
(solid line with circular symbols); and gravitational evolution with frozen
hydrodynamics using quasi-Newtonian data (dotted lines with triangular
symbols). After some transient behavior lasting until $u\sim 2$, all curves
roughly approach the same value.}

\label{schw_comp}
\end{figure}

As an illustration of this behavior we evolve a star in a space-time both with
and without a black hole. The star is placed at $a=9$ with mass $m=10^{-4}$
and radius $R_*=3$ and the gravitational initial data is given either by $J=0$
(shear-free data) or by the quasi-Newtonian value. For purposes of comparison, runs are made both
with the hydrodynamic variables evolved and frozen. Figures
\ref{mink_comp} and \ref{schw_comp} illustrate the behavior of $J$ for these
cases. After a relatively brief transient behavior, the norm of $J$ approaches
a value quite independent of the initial data  (with similar behavior observed
in all other gravitational field variables). This indicates that most
of the spurious radiation present in the initial hypersurface is `flushed out'
in a short time. In order to elucidate whether the system settles into a
quasi-stationary state, as might be expected, we analyze next whether there is
an approximate helical Killing vector in the space-time.

\subsection*{{\bf Agreement of the ``relaxed state''}} 
\begin{figure}
\begin{center}
\epsfig{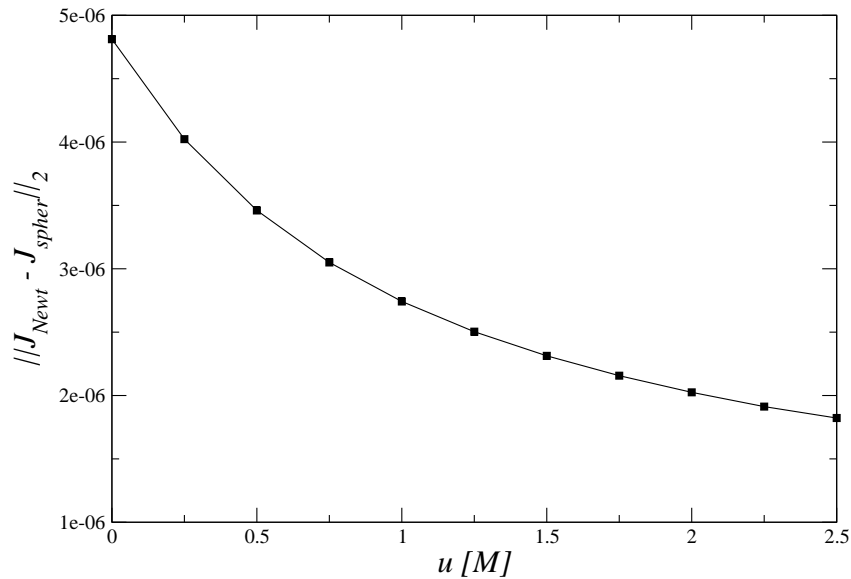}
\end{center}
\caption{$L_2$ norm of the difference between the
computed values of $J$ for
the quasi-Newtonian and spherically symmetric initial data.}
\label{difference}
\end{figure}

\begin{figure}
\begin{center}
\epsfig{file=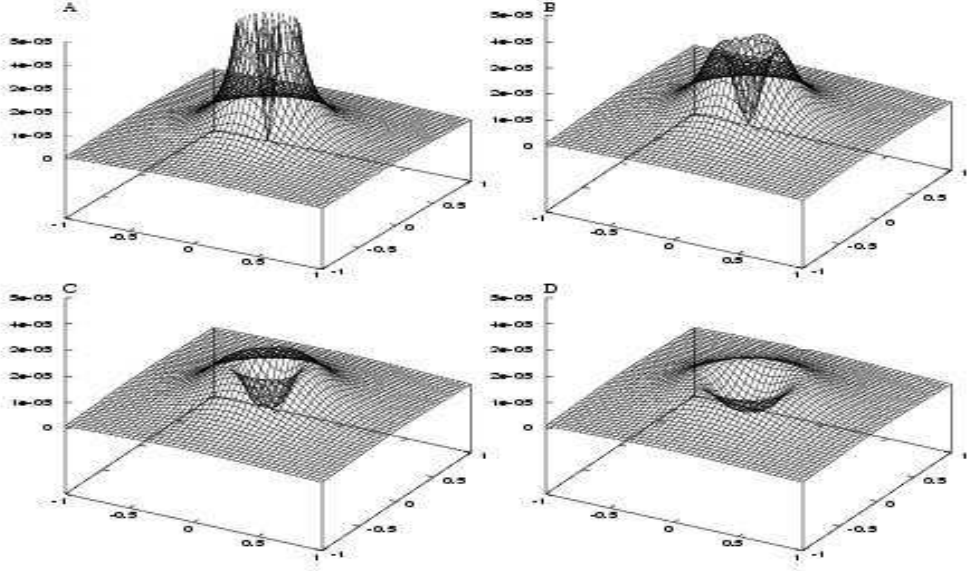,height=3.5in,width=5.5in,angle=0}
\end{center}
\caption{Absolute value of the difference between the computed values of $J$ at
${\cal I}^+$ evolved from quasi-Newtonian and from shear-free initial data. The
panels (A,B,C,D) correspond to the time sequence $u=0$, $u=0.75M$, $u=1.5M$, and $u=2.5M$. Panel A has been truncated for comparison purposes, as
its height is $\approx 9 \times 10^{-5}$.}
\label{difference2}
\end{figure}

As indicated above, the
system appears to relax to a state independent of the details of the initial
data. In order to quantify this,  we calculate the $L_2$ norm of the difference
between the two numerical solutions obtained with Newtonian and with shear-free
initial data. Figures~\ref{difference} and \ref{difference2} illustrate the
behavior of the difference vs. time, showing how as time progresses the
agreement between the numerical solutions becomes more pronounced.

It is also illustrative to observe the point-wise difference between the
two numerical solutions. In particular, inspection of  $|J_{Newt}-J_{spher}|$
at ${\cal I}^+$ reveals not only that the solutions tend to agree as time
progresses, but also that this agreement is more marked along null rays passing
through the star and neighboring rays.

\subsection{{\bf Quasi-equilibrium behavior}}

In order to measure the approach to quasi-equilibrium we monitor the
rate of change of the angular part of the metric along the streamlines of the
vector field  $\xi = T + \kappa \Omega \Phi$. As explained in Sec.~\ref{sec:U},
for $\kappa=1$, $\xi$ equals the helical Killing vector of the background black
hole with $\Omega$ set to the initial orbital angular velocity of the star. For
comparison purposes we also consider $\kappa=0$, for which $\xi$ equals the
static Killing vector $T$ of the background black hole, and  $\kappa=-1$, for
which $\xi$ equals the helical Killing vector counter-rotating with respect to
the orbital motion of the star. We measure the change of the gravitational field
with respect to the flow of these vector fields by the norm $F_\kappa \equiv
||{\cal L}_{\xi} h_{AB}||^2$. For orbital motion in quasi-equilibrium around the
black hole, we should then find $F_{\kappa=1}\approx 0$. In order to verify that
this is the case, we monitor the values of $F_\kappa$  for $\kappa=1,0,-1$.
Again, the star is initialized at $a=9$ with mass $m=10^{-5}$ and radius
$R_*=3$.

\begin{figure}
\begin{center}
\epsfig{file=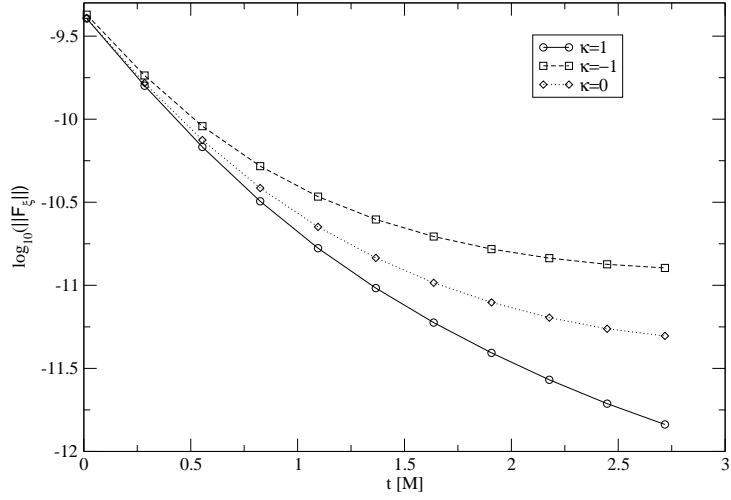,height=3.in,angle=0}
\end{center}
\caption{Comparison of $F_\kappa$ using quasi-Newtonian initial data.}
\label{killing1}
\end{figure}

The first test, carried out with quasi-Newtonian initial data and shown in Figure~\ref{killing1}, compares the behavior of $F_\kappa$ for the three different
values of $\kappa$ indicated above. At first, the curves differ very little, but at later times
there is a marked difference ($>1.5$ orders of magnitude), indicating that the
system approaches an approximate helical symmetry. For the $\kappa=1$ helical
case, $F_1$ decays more than two orders of magnitude from its initial
value.

\begin{figure}
\begin{center}
\epsfig{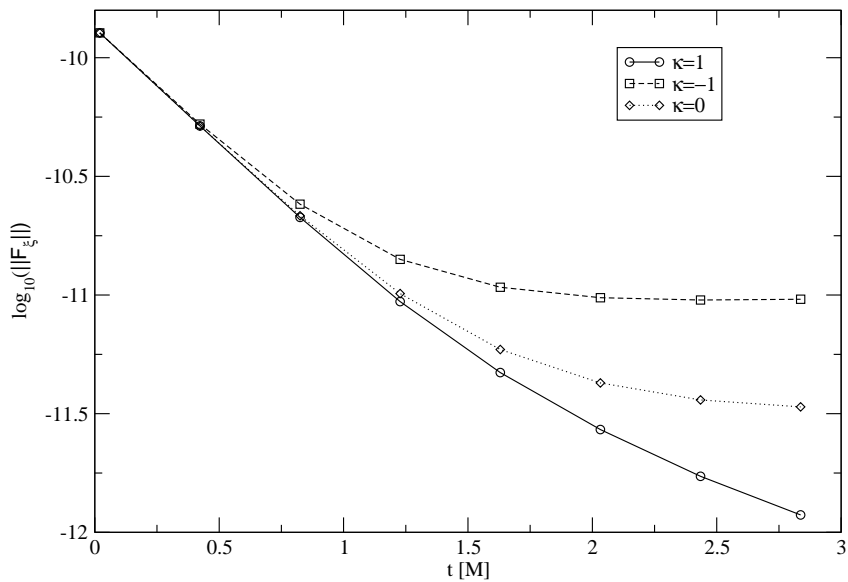}
\end{center}
\caption{Comparison of $F_\kappa$  using shear-free initial data.}
\label{killing2}
\end{figure}

Next we repeat the test with $J=0$ shear-free initial data, with the results shown in Fig.~\ref{killing2}.
The initial values
of $F_\kappa$ are
not the same as those
in the first test, but a
similar time behavior is observed. Again the helical choice $F_1$
decays more than two orders of magnitude from its initial value,
and more than an order of magnitude more than either $F_0$ or
$F_{-1}$.

\subsection{Co-rotating coordinates}

\begin{figure}
\begin{center}
\epsfig{file=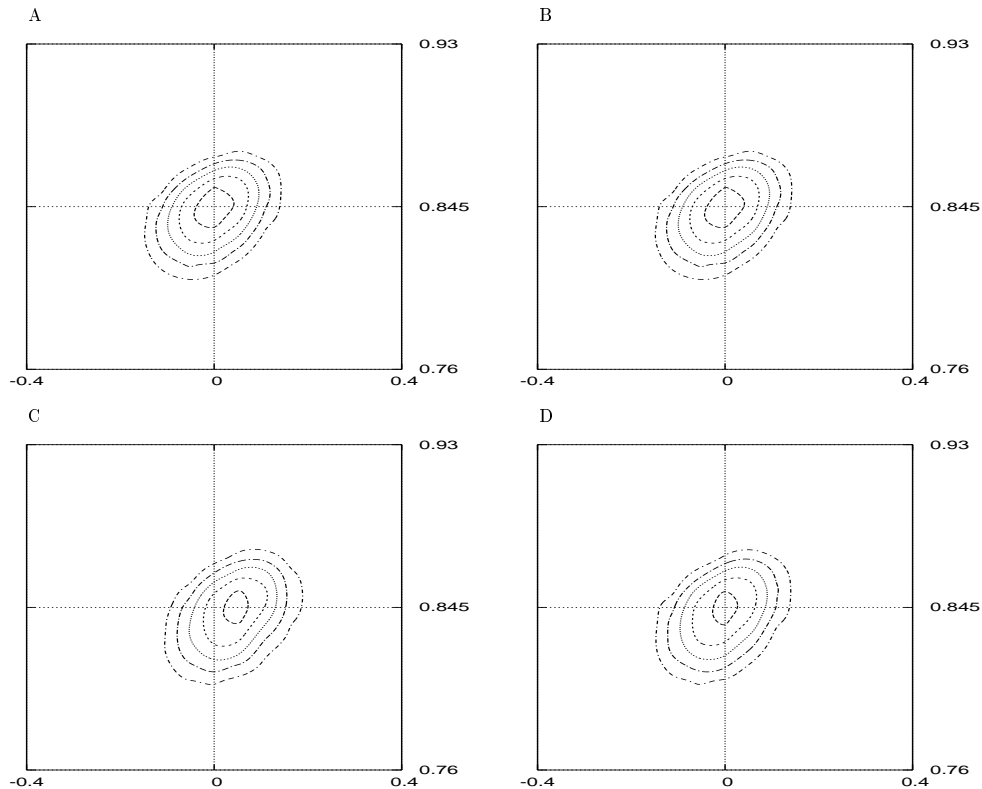,height=4.5in,angle=0}
\end{center}
\caption{For the case $m=10^{-5}$, $R_*=3M$, $a=9M$ we show density profiles in
the plane with stereographic coordinate $p=0$. The motion takes place in the
$(r,q)$ plane. In all figures, the vertical axis is $x$ (the compactified radial
coordinate)and the horizontal axis is $q$. The top panels correspond to the fixed
coordinates (A) and co-rotating coordinates (B) at the initial time $u=0$. Figures
(C) and (D) show the corresponding density profiles at $u=2.8$. The co-rotating
coordinates, panel (D), perform well in keeping the coordinate location of the
star fixed. For comparison, panel (C) shows the actual displacement of the star.}
\label{figrotnorot}
\end{figure}

Co-rotating coordinates can improve the tracking and hydrodynamic treatment of
the orbiting star by reducing or eliminating the variation of field values due to purely coordinate effects. They are introduced here by specifying the value of $U$ at the
inner boundary, as  described in section~\ref{sec:U}. By defining $U$ via
Eq.~(\ref{corotation}), we have checked that this indeed keeps the angular
coordinate of the star fixed. The tests were carried out in the star mass range $m\in
[10^{-7},10^{-2}]$, with $M=1$, $R_*=3M$ and $a=9M$. Figure \ref{figrotnorot}
illustrates the results for $m=10^{-5}$. As can be seen clearly from the density contours displayed, the co-rotating coordinates maintain
the central density of the star at the same initial coordinate location after evolution
through $u=2.8$. Note that, although the initial polytrope is spherically
symmetric about the center of the star, the kinetic energy contributed by the
initially uniform orbital angular velocity skews the density distribution. These
results verify that the initialization can be performed equally well in
co-rotating coordinates.

\subsection{{\bf Monitoring the expansion of the null coordinates}}
\label{sec:mexpans}

\begin{figure}
\begin{center}
\epsfig{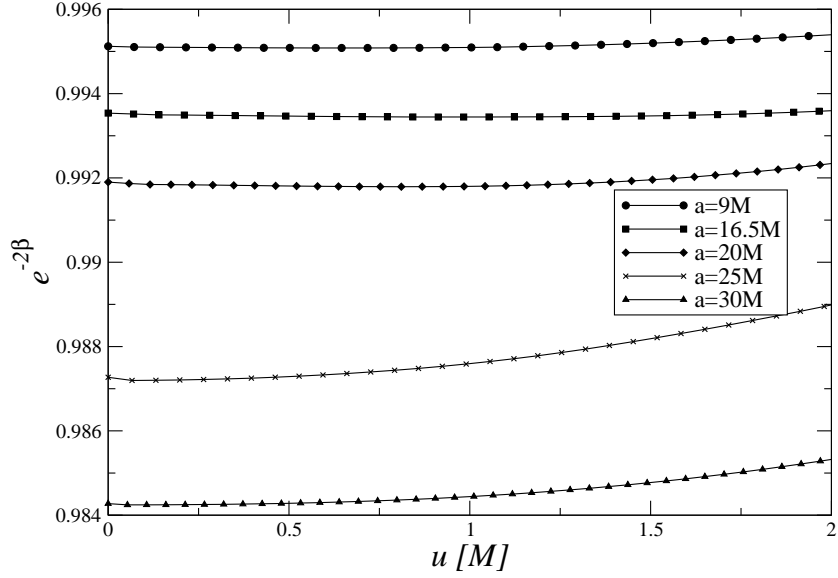}
\end{center}
\caption{Time behavior of the minimum expansion $e^{2\beta}$ at future null
infinity vs. initial location of the star. As the star is placed farther from
the black hole its focusing effect on the central null rays increases. However,
the expansion stays close to its initial value for a given separation $a$.}
\label{beta1}
\end{figure}

When the star is initialized sufficiently far from the central black hole, its
bending effect on neighboring light rays can cause a coordinate singularity, by
reducing the expansion of the outgoing null hypersurfaces to zero. As discussed
in Sec.~\ref{sec:caustic}, the bending effect on outgoing null rays is first
evident at future null infinity, where it is manifested in our formalism by
$\beta\rightarrow \infty$. However, $\beta\rightarrow \infty$ also when the
generators of the outgoing null hypersurfaces approach a black hole horizon, in
which case $\beta\rightarrow \infty$ on a complete spherical set of rays. As a
result, it is difficult to distinguish at a given retarded time between whether
it is a horizon or a coordinate singularity that is responsible for the lack of
expansion. In the present context, what matters is whether such coordinate singularities can arise on a relatively
short dynamical time scale which would have bearing on our main conclusion that
the system relaxes to a quasi-stationary orbit.
To assess this possibility, we carry out evolutions with a star of mass $m=10^{-3}M$
and radius $R_*=3M$ initially placed in orbit at different distances $a$
from the central black hole. We then
monitor the expansion $e^{-2\beta}$ for roughly the ``relaxation'' time.
Figure~\ref{beta1} graphs the $u$-dependence of the minimum value of
$e^{-2\beta}$ over the sphere at null infinity for the  initial orbital radii
$a=9M,16.5M, 20M, 25M, 30M$. As expected, the expansion becomes smaller as the
separation between the star and the black hole increases. More important to the
initialization problem, for a given separation the minimum expansion does not
appear to change considerably from its initial value during the relaxation time.
An estimate obtained with the lens equation combined with the initial value of
$\beta$ at infinity seems to provide a good indicator of whether coordinate
singularities will develop.

\begin{figure}
\begin{center}
\epsfig{file=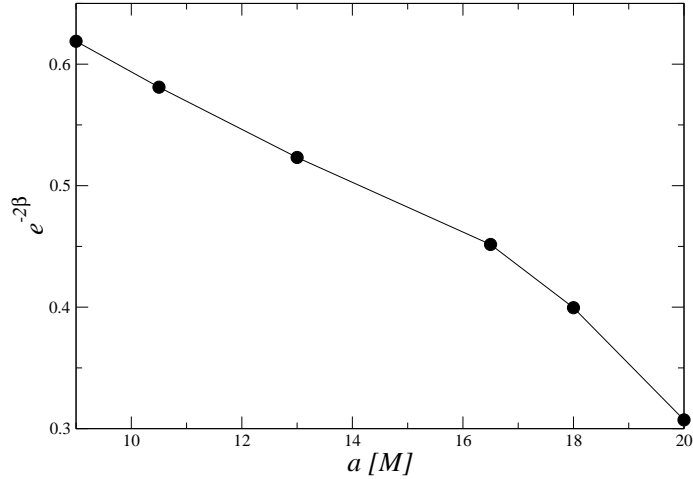,height=3in,angle=0}
\end{center}
\caption{Behavior of the minimum of the expansion $e^{-2\beta}$ at
future null infinity vs. location of the star. As the star's
initial location is placed farther from the black hole its
focusing effect on the central null rays becomes considerably.}
\label{beta2}
\end{figure}

We now examine the possibility of dealing with a more massive star
$m=10^{-1}M$, keeping $R_*=3M$. The lens estimate predicts zero expansion at a
separation given by $a\ge 22.5M$. To study this problem, we set up data at
different separations and, in view of the behavior just seen in the case with
$m=10^{-3}M$, we evolve for a short time ($u=M/10$).  Figure~\ref{beta2} graphs
the computed value of $e^{-2\beta}$ vs. $a$, for a range of separations. As the
star is placed farther from the black hole, the expansion gets considerably
smaller. Furthermore, the values obtained for $a>16M$  only represent an upper
bound since even with the largest practical grid $nr \times nq \times np = 132
\times 82^2$ the star is considerably unresolved. As a result, the neighboring
null rays pass by the star at a significant ``grid distance''  and do not
accurately represent the expansion that would be calculated in the continuum
problem. Nevertheless, the results in the figure suggest that it would be
possible to simulate a star in orbit around a black hole with $a < 16M$ on a
uniform grid without developing coordinate problems.

\subsection{Code performance}

We illustrate the speed of the code for the case of the finest
grid used in the convergence test, i.e. $81^2 \times 123$. Setting the time-step to $\Delta u = 0.014M$, 
a run until $u=1.5M$ takes 9 hours on a 2.4GHz Pentium 4  processor,
and requires 1.4Gb of memory. From this we conclude that a single orbit would take roughly 1 1/2 months.

\section{Conclusion}
\label{sec:conc}

Within the characteristic framework, we have developed and implemented a
numerical relativity code, as well as procedures for finding the required
initial data, for evolving a star in close orbit around a Schwarzschild black
hole. We have shown that after a short evolution time the system relaxes to a
quasi-equilibrium state which is mainly independent of the initial {\it
gravitational} data.  Variations of the initial matter data, such as the shape
or size of the star, have not been investigated. For small variations about
stellar equilibrium one would expect similar results since the accompanying
variations in the star's gravitational field should relax to the same
quasi-equilibrium state. We have also developed and demonstrated tools that
permit the use of co-rotating coordinates and that monitor possible problems
with the null coordinate system.

Successful simulation of a neutron star in close orbit around a black hole
requires a gravity code, a hydrodynamic code and physically appropriate initial
data. It has already been demonstrated that the characteristic gravity code is
accurate and stable~\cite{hpn,roberto}. In this paper we have implemented and
tested a conservative formulation of the hydrodynamics and used it to shed
light on the problem of prescribing physically meaningful initial data. There
are several issues that remain to be addressed before physically interesting
long term evolutions can be carried out. The results obtained in this paper
would appear to justify the effort required to make these improvements.

Specifically, in order for the code to yield astrophysically useful results,
three further conditions must be fulfilled. First, more realistic inner
boundary data must be provided. This must correspond to a spinning black hole
and must include the gravitational distortion induced by the companion star.
Second, the turn-around of results must be considerably improved. This will
require revising the numerical algorithms and parallelization of the code to
take advantage of large platforms. Finally, since accurate simulations require
that the numerical error be well below the expected radiation output ($<
5$\%), the use of adaptive mesh refinement seems crucial. Preliminary work in
this direction has recently been presented~\cite{nullamr}. All of these items
are major tasks and are deferred to future work.

\begin{acknowledgments}
We wish to thank Matthew Choptuik, Dave Neilsen, Mark Miller, Lee Lindblom and
Jorge Pullin for discussions and Erik Schnetter for comments on the manuscript. 
We benefited from the hospitality of the
Max-Planck-Institut f\"ur GravitationsPhysik, Albert-Einstein-Instit\"ut.
N.T.B. thanks Louisiana State University for hospitality. The work  was
supported by in parts by the National Research Foundation, South Africa under
Grant number 2053724, the NSF under grants NSF-0242507 and 0244699 to
Louisiana State University and the Horace Hearne Lab of Theoretical Physics.
L.L. was supported in part by an Alfred P. Sloan Fellowships. J.W.
acknowledges research support under National Science Foundation Grant
PHY-0244673 to the University of Pittsburgh.
R.G. acknowledges partial support under National Science Foundation Grant
PHY-0135390 to Carnegie Mellon University.
\end{acknowledgments}

\appendix

\section{Computation of the quasi-Newtonian initial data}

The Newtonian potential outside the polytrope is simply
\begin{equation}
\Phi=-\frac{m}{R} \; \; \; (R>R_*).
\end{equation}
Inside the polytrope, the derivation is rather more complicated.  We find
\begin{equation}
\Phi=-\frac{m}{R_*}-\frac{m \sin{\frac{R \pi}{R_*} } }{R \pi}
         \; \; \; (R<R_*).
\end{equation}
and from it obtained $\eth^2 \Phi$. First, we make the following
definitions
\begin{eqnarray}
z & = & q+ip \nonumber \\
t_1 & = & (r-a)^2 + z \bar{z} (r+a)^2 \nonumber \\
P_p & = & 1 + z \bar{z} \nonumber \\
t_2 & = & \frac{\pi}{R_*} \sqrt{\frac{t_1}{P_p}}  \nonumber \\
t_3 & = & \frac{a^2 r^2 z^2 m \sqrt{P_p}}{t_1^{5/2}}.
\end{eqnarray}
Then we may write
\begin{eqnarray}
\eth^2 \Phi & = & -12 t_3 \; \; \; (R>R_*) \nonumber \\
\eth^2 \Phi & = & -4 \frac{t_3}{R_*^2 \pi} \times \nonumber \\
& & \left(
( -\pi^2 r^2 +2 r \pi^2 a \frac{1-z \bar{z}}{P_p} + 3 R_*^2
-\pi^2 a^2)\sin{t_2}
-3  R_* \pi \sqrt{\frac{t_1}{P_p}} \cos{t_2} \right) \nonumber \\
& & (R<R_*).
\label{eq:eth2}
\end{eqnarray}
Finally, $J$ is obtained by numerically integrating $(r^2 J_{,r})_{,r} = - 2 \eth^2 \Phi$
in second order form starting from the inner boundary $r=2M$.


\begin{thebibliography}{99}

\bibitem{rees}
M.~J.~Rees,
in {\it Proceedings of the 18th Texas Symposium},
edited by A. Olinto, J. Frieman and D. Schramm
(World Scientific, Singapore, 1998).

\bibitem{ct}
C. Cutler and K.S. Thorne, in
{\it Proceedings of the 16th International Conference on General
Relativity \& Gravitation}, edited by
N.T. Bishop and S.D. Maharaj (World Scientific, Singapore, 2002).

\bibitem{vallisneri}
M.~Vallisneri,
Phys.\ Rev.\ Lett.\  {\bf 84}, 3519 (2000).

\bibitem{davies}
M.~B.~Davies, A.~J.~Levan and A.~R.~King,
``The ultimate outcome of black hole - neutron star mergers'',
Mon. Not. R. Astron. Soc, to appear,
astro-ph/0409681.

\bibitem{baumgarte}
T. Baumgarte, private communication.

\bibitem{shibata}
M. Shibata, private communication.

\bibitem{miller}
M. Miller, private communication.

\bibitem{hawke}
I. Hawke, private communication.


\bibitem{siebel1}
F.~Siebel, J.~A.~Font, E.~Muller and P.~Papadopoulos,
Phys. Rev. D {\bf 67}, 124018 (2003).

\bibitem{siebel2}
F.~Siebel, J.~A.~Font and P.~Papadopoulos,
Phys.\ Rev.\ D {\bf 65}, 024021 (2002).

\bibitem{hpn}
N.T. Bishop, R. G\'omez, L. Lehner, M. Maharaj, and J. Winicour, Phys. Rev. D
{\bf 56}, 6298 (1997).

\bibitem{mat}
N.T. Bishop, R. G\'omez, L. Lehner, M. Maharaj, and J. Winicour, Phys. Rev. D
{\bf 60}, 024005 (1999).

\bibitem{klein}
C.~Klein,
Binary black hole spacetimes with a helical Killing vector,
Phys. Rev. D, to appear,
gr-qc/0410095.

\bibitem{shibatafriedman}
M.~Shibata, K.~Uryu and J.~L.~Friedman,
Phys.\ Rev.\ D {\bf 70}, 044044 (2004).

\bibitem{yocook}
H.~J.~Yo, J.~N.~Cook, S.~L.~Shapiro, and T.~W.~Baumgarte,
Phys. Rev. D {\bf 70}, 084033 (2004).

\bibitem{andrade}
Z.~Andrade et. al.,
Phys. Rev. D {\bf 70}, 064001 (2004).

\bibitem{grandclement}
E.~Gourgoulhon, P.~Grandcl\'ement and S.~Bonazzola,
Phys.\ Rev.\ D {\bf 65}, 044020 (2002).

\bibitem{leveque} R.~J. Leveque,
{\it Numerical Methods for Conservation Laws} (Birkhauser-Verlag, Basel, 1990).

\bibitem{philiptoni}
P.~Papadopoulos and J.~A.~Font,
Phys.\ Rev.\ D {\bf 61}, 024015 (2000).

\bibitem{davis}
S. Davis, Siam J. Sci. Stat. Comput., {\bf 8}, 1, (1987).

\bibitem{newt1}
J. Winicour, J. Math. Phys. {\bf 24}, 1193 (1983).

\bibitem{newt2}
J. Winicour, J. Math. Phys. {\bf 24}, 2506 (1983).

\bibitem{iww}
R.A. Isaacson, J.S. Welling, and J. Winicour, J. Math. Phys. {\bf 26},
2859 (1985).

\bibitem{qf}
J. Winicour, J. Math. Phys. {\bf 28}, 668 (1987).

\bibitem{cce}
N.T. Bishop, R. G\'omez, L. Lehner, and J. Winicour, Phys. Rev. D {\bf 54}, 6153
(1996).

\bibitem{rai83}
R.A. Isaacson, J.S. Welling, and J. Winicour, J. Math. Phys. {\bf 24}, 1824
(1983).

\bibitem{ntb93}
N.T. Bishop, Class. Quantum Grav. {\bf 10}, 333 (1993).

\bibitem{ntb90}
N.T. Bishop C.J.S. Clarke, and R.A. d'Inverno, Class. Quantum Grav. {\bf 7}, L23
(1993).

\bibitem{bondi}
H. Bondi, M.J.G. van der Burg and A.W.K. Metzner,
{\em Proc. R. Soc. A} {\bf 269} 21, (1962).

\bibitem{sachs}
R.K. Sachs, Proc. R. Soc. London A {\bf 270},  103  (1962).

\bibitem{tam}
L.A. Tamburino and J. Winicour, Phys. Rev. {\bf 150}, 1039, (1966).

\bibitem{roberto}
R. G\'omez, Phys. Rev. D {\bf 64}, 024007 (2001).

\bibitem{particle}
N.T. Bishop, R. G\'omez, S. Husa, L. Lehner, and J. Winicour, Phys. Rev.
D {\bf 68}, 084015 (2003)

\bibitem{modemode}
Y.~Zlochower, R.~G\'omez, S.~Husa, L.~Lehner and J.~Winicour,
Phys.\ Rev.\ D {\bf 68}, 084014 (2003).

\bibitem{eth}
R. G\'omez, L. Lehner, P. Papadopoulos and J. Winicour, Class. Quantum
Grav. {\bf 14}, 977, (1997).

\bibitem{chandra1}
S. Chandrasekhar, {\it An introduction to the study of stellar structure}
(Dover, New York, 1967).

\bibitem{uryu}
K. Uryu and Y. Eriguchi, Mon. Not. R. Astron. Soc. {\bf 303}, 329 (1999).

\bibitem{lom}
J.C. Lombardi, F.A. Rasio, S.L. Shapiro, Phys. Rev. D {\bf 56}, 3416 (1997).

\bibitem{nulltube}
R. G\'omez, S. Husa and
J. Winicour, Phys. Rev. D {\bf 64}, 0240010 (2001).

\bibitem{wobble}
R.~G\'omez, L.~Lehner, R.~L.~Marsa and J.~Winicour,
Phys. Rev. D {\bf 57}, 4778 (1998).

\bibitem{fission}
R.~G\'omez, S.~Husa, L.~Lehner and J.~Winicour,
Phys. Rev. D {\bf 66}, 064019 (2002).

\bibitem{nullamr}
F.~Pretorius and L.~Lehner,
J. Comput. Phys. {\bf 198}, 10 (2004).

\bibitem{dain}
S.~Dain,
``A new geometric invariant on initial data for Einstein equations'', preprint,
arXiv:gr-qc/0406099.

\end{thebibliography}
\end{document}